\documentclass[twoside]{ilcws10}
\usepackage[latin1]{inputenc}
\usepackage[dvips]{graphicx,epsfig,color}
\usepackage{wrapfig,rotating}
\usepackage{amssymb,amsmath,array}

\def\zh2qqcc{$ZH \to q\bar{q}c\bar{c}$}
\def\qqqq{q\bar{q}q\bar{q}}
\def\qq{q\bar{q}}
\def\nn{\nu\bar{\nu}}
\def\ll{{\ell}{\ell}}
\def\llll{{\ell}{\ell}{\ell}{\ell}}
\def\nlqq{{\nu}{\ell}{q}{\bar{q}}}
\def\nnqq{{\nu}{\bar{\nu}}{q}{\bar{q}}}
\def\llqq{{\ell}{\ell}{q}{\bar{q}}}

\begin{document}

\title{\bf Branching ratio study of $ZH\to q\bar{q}c\bar{c}/q\bar{q}b\bar{b}$ }

\author{Hiroaki Ono$^1$
\thanks{TEL:+81-25-267-1500 (537), MAIL: ono@ngt.ndu.ac.jp}
\vspace{.3cm}\\
1 - Nippon Dental University School of Life Dentistry at Niigata\\
1-8 Hamaura-cho chuo-ku Niigata, Niigata - Japan
}

\maketitle

\begin{abstract}
 Precise measurement of the Higgs boson properties is an important issue
 of the International Linear Collider (ILC) experiment
 to verify the particles mass generation mechanism that
 the coupling strength between the Higgs boson and the fermions or vector bosons
 are proportional to the mass of each particle.
 Thus the measurement of the branching ratio of the Higgs boson an important issue
 to understand the mass of each particle.
 In this analysis,
 measurement accuracy of the Higgs boson branching ratio
 in the $ZH \to q\bar{q}H$ hadronic decay mode was studied
 with the cut-based analysis in Higgs mass of $M_{H}=120~{\rm GeV}$
 at the center-of-mass energy of $\sqrt{s}=250~{\rm GeV}$ with the ILD detector model.
 From the analysis, we estimate the measurement accuracy of the relative Higgs boson branching ratio
 of $BR(H \to c\bar{c})$ to $BR(H \to b\bar{b})$ as 13.68\%.
\end{abstract}
 \begin{wrapfigure}{rt}{0.4\columnwidth}
  \begin{center}
   \includegraphics[width=0.4\textwidth]{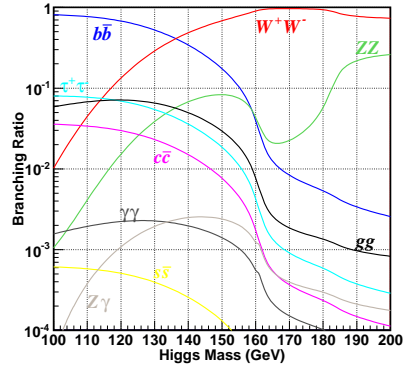}
   \caption{Branching fraction of the Higgs boson decay at the lower mass range ($100 \leq M_{H} \leq 200~{\rm GeV}$).}
   \label{fig:ZH_branch}
  \end{center}
 \end{wrapfigure}

 \section{Introduction}
 
\begin{wrapfigure}{l}{0.5\textwidth}
  \begin{center}
   \includegraphics[width=0.4\textwidth]{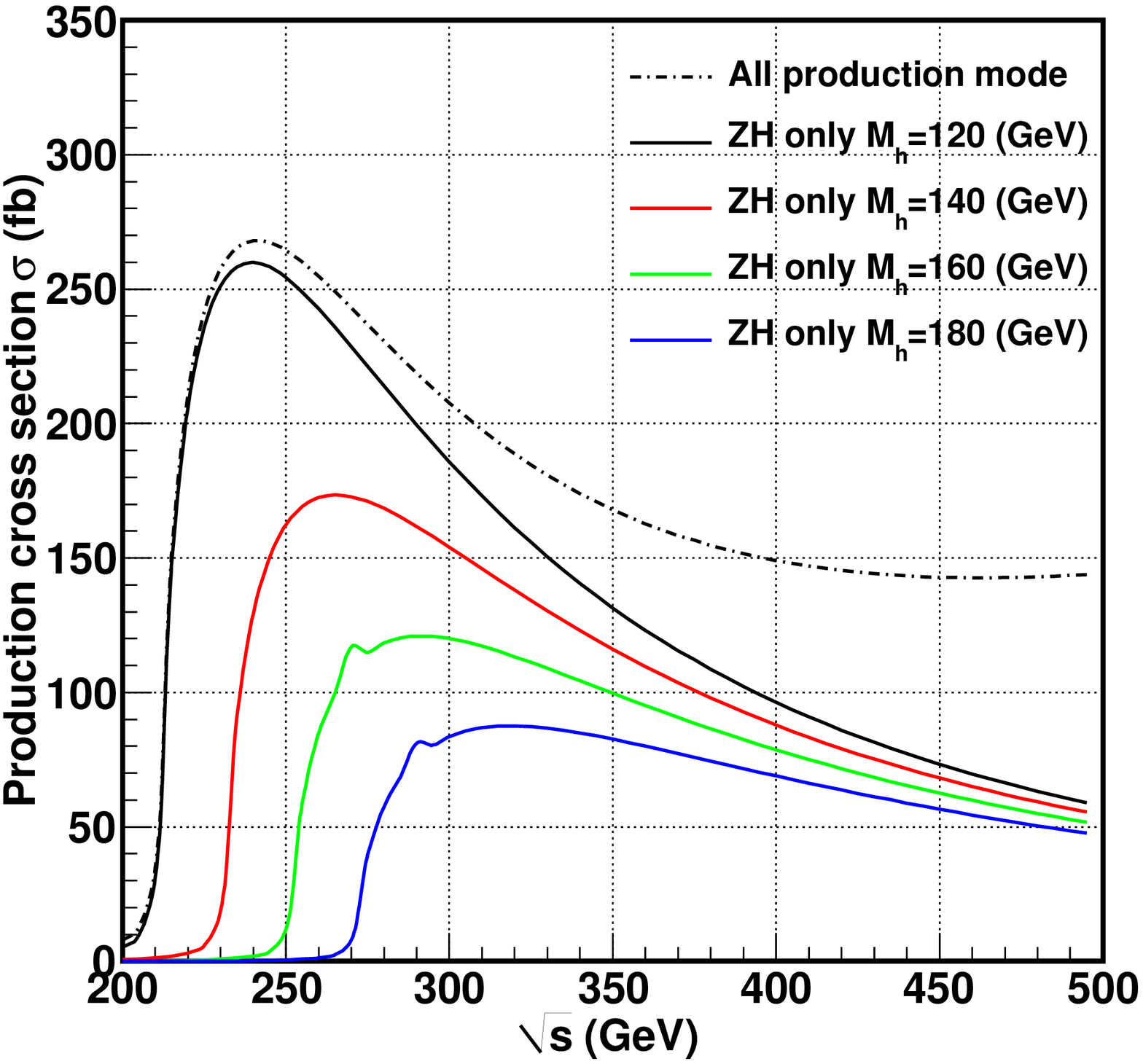}
   \caption{Production cross section of the Higgs with $\sqrt{s}$}
   \label{fig:ZH_xsec}
  \includegraphics[width=0.4\textwidth]{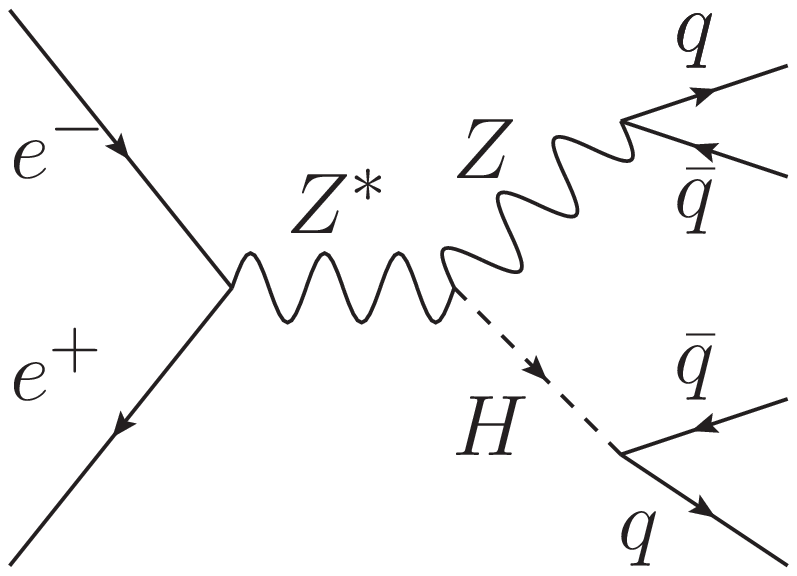}
  \caption{Higgs boson production via Higgs-strahlung $(ZH)$ process.}   
  \label{fig:ZH_diagram}
  \end{center}
 \end{wrapfigure}

  International Linear Collider (ILC)~\cite{RDR} is a future $e^{+}e^{-}$ collider experiment
 for the precise measurement and the validation of the Standard Model (SM) physics,
 especially for the measurement of the Higgs boson property,
 even the discovery of the Higgs boson will be realized in Large Hadron Collider (LHC) experiment.
 In the SM, light Higgs boson mass ($M_{H}$) is predicted the range of
 $M_{H} \geq 114.4~{\rm GeV}$ from the study in Large Electron Positron Collider (LEP II)~\cite{LEPEWWG}
 and recently Tevatron experiments exclude the Higgs mass range around
 $160 \leq M_{H} \leq 170~{\rm GeV}$ with the 95\% confidence level~\cite{CDF}.
 From these results,
 Higgs mass is indicated to be light ($M_{H} \leq 140~{\rm GeV}$)
 and in this region, Higgs mainly decays to $b\bar{b}$ pair which forms multi-jet final state,
 as shown in Fig.~\ref{fig:ZH_branch}.
 Since hadron collider experiments have large QCD multi-jet backgrounds,
 the measurement of the light mass Higgs
 will not prefer in terms of the signal to noise ratio, 
 ILC experiment has an advantage for precise measurement of light Higgs boson
 with large signal yield of multi-jet final state in lower background environment.
 Therefore, precise measurement of the light mass Higgs ($M_{H} \leq 140~{GeV}$)
 will be a primary target of the ILC experiment.
 At the lower center-of-mass energy ($\sqrt{s}$) range around the production threshold, as shown in Fig~\ref{fig:ZH_xsec},
 SM Higgs boson is mainly produced through the Higgs-strahlung ($ZH$) process,
 which associated with the $Z$ boson as shown in Fig.~\ref{fig:ZH_diagram}.

 The largest production cross-section via $ZH$ mode is obtained
 around the center-of-mass energy ($\sqrt{s}$)
 at the $ZH$ production threshold region as shown in Fig.~\ref{fig:ZH_xsec} (a).
 Since the $Z$ boson mainly decays to $q\bar{q}$ pair,
 the largest Higgs boson production cross-section via $ZH$ process is obtained
 through the $ZH \to q\bar{q}H$ process.
 Therefore, we study the Higgs boson property
 with the largest production cross-section process of $ZH \to q\bar{q}H$.
 Since Higgs boson mainly decays to $b\bar{b}$ pair at the Higgs mass below 140~GeV region
 as shown in Fig.~\ref{fig:ZH_xsec} (b),
 the final state of the $ZH \to q\bar{q}H$ process forms the four-jet.
 In ILC experiment,
 two detector concepts, ILD and SiD submit their Letter of Intent (LOI) and
 validated by ILC Detector Advisory Group (IDAG).
 In order to achieve the best jet energy resolution in multi-jet environment,
 ILD adopt the Particle Flow Algorithm (PFA) suited detector design,
 which has fine-segmented calorimeter with strong magnetic field.
 In this analysis,
 we study the measurement accuracy of the branching ratio of Higgs boson
 with the full detector simulation for $ZH \to q\bar{q}H$ hadronic mode with the ILD detector model.

 \section{Analysis tools and MC samples}

 \begin{wraptable}{r}{0.55\columnwidth}
  \begin{center}
   \caption{Signal and background data samples.}
   \begin{tabular}{|l|l|}
    \hline
    Signal and Bkg & DST data samples\\
    \hline
    Higgs sample & $\qq H$, $\nn H$, $\ll H$\\
    \hline
    SM background & $\qqqq, \nlqq, \nnqq, \qq, \llll, gg$\\
    \hline
   \end{tabular}\\
  \end{center}
 \end{wraptable}

 For full detector simulation study,
 we use the ILD detector model based Monte Carlo (MC) full simulation package called Mokka,
 which is based on the MC simulation package Geant4~\cite{Geant4}.
 Generated MC hits are reconstructed and smeared in the reconstruction package called MarineReco
 which includes the PFA package called PandoraPFA~\cite{ilcsoft}.
 Reconstructed his are skimmed and saved in the ILC common data format called LCIO.
 For the event analysis,
 we use the useful analysis package library called Anlib
 and each analysis procedure is handled through JSF~\cite{SimTools} based on Root~\cite{Root}.
 In this analysis, we assume the center-of-mass energy around the $ZH$ production threshold of
 $\sqrt{s}=250~{\rm GeV}$ and the light Higgs mass of $M_{H} = 120~{\rm GeV}$.
 Each data sample is scaled to the integrated luminosity of $\mathcal{L} =250~{\rm fb^{-1}}$
 and the beam polarization to $P(e^{+}, e^{-})=(+30\%, -80\%)$.
 The main backgrounds for $ZH$ hadronic decay mode are considered as following processes:
 $ZH \to Z^{*}/\gamma \to q\bar{q}$,
 $e^{+}e^{-} \to WW/ZZ \to qq'q''q'''~{\rm or}~q\bar{q}q'\bar{q'},~\llqq,~\nnqq,$
 $e^{+}e^{-} \to WW \to {\nu}{\ell}qq'$ and $e^{+}e^{-} \to ZZ \to \llll$.

 \section{Event reconstruction}
 \begin{wrapfigure}{r}{0.5\columnwidth}
  \begin{center}
   \includegraphics[width=0.4\textwidth]{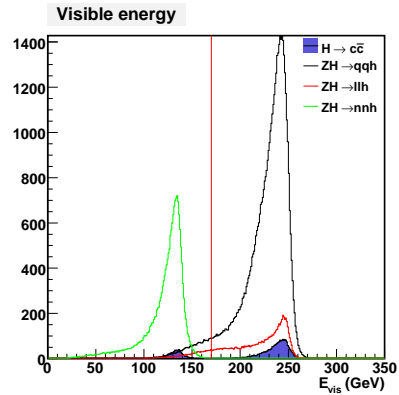}
   \caption{Visible energy distribution to select the $Z\to\qq$ hadronic mode from pre-mixed sample.}
  \end{center}
 \end{wrapfigure}
  
  Since the final state of the $ZH \to q\bar{q}H$ mode forms four-jet,
 after the PandoraPFA clustering,
 forced four-jet clustering based on Durham jet-clustering algorithm has applied.
 In order to select the best jet pair combination from the four-jet,
 $d=\left( M_{12}-M_{Z} \right)^2+\left( M_{34}-M_{H} \right)^2$ value is evaluated,
 where $M_{12}$ is a $Z$ candidate di-jet mass,
 $M_{34}$ is a Higgs candidate di-jet mass,
 $M_{Z/H}$ are $Z$ and Higgs boson masses.
 From the four-jet, minimum $d$ value jets combination is selected as best $Z$ and $H$ candidates.
 In order to improve the background reduction from $WW/ZZ$ with mass distribution,
 kinematic fitting is applied with following constraints after the jet pairing;
 $\sum E_{i} - E_{CM}=0$, $\sum \vec{P_{i}} = 0$, $|M_{12}-M_{34}| = |M_{Z}-M_{H}|$
 (di-jet mass difference consistent with $Z$ and $H$ mass difference),
 where $E_{i}$ and $\vec{P_{i}}$ are $i$-th jet energy and momentum which is sorted by energy, respectively.\\
 After the jet pair combination and kinematic fitting,
 event selections are applied.
 For the $ZH$ hadronic decay mode sample ($ZH \to \qq H$) selection from the pre-mixed sample of
 $ZH\to \qq/\nn/\ll H$, following signal classification is applied:
 (0). Visible energy cut : $E_{vis} > 170~{\rm GeV}$ and no high momentum tracks ($P_{lepton}>15~{\rm GeV}$),
 After the classification, we apply the following selection criteria.
 In order to select the four jet reconstructed events,
 we require the number of charged tracks in each jet above four ($N_{charged} > 4$)
 and logarithm of the jet reconstruction $Y$-value threshold from three to four jet ($Y_{34}$)
 should be $-\log{Y_{34}} > 2.7$.
 After that, thrust, thrust angle ($\cos{\theta_{thrust}}$) and jets angle ($\theta_{H, Z}$) cuts
 are required to suppress the $ZZ$ background from the difference of sphericity of final state jet shape.
 Finally we require the consistency of mass of the reconstructed jets pair after the kinematic fit
 that $Z$ candidate jets pair should be consistent with $M_{Z}$ and
 the other pair should be consistent with $M_{H}$.
 For the reduction of hard photons and ISR photons, finally highest photon energy cut is applied.
 Table.~\ref{cut:table} shows the background reduction summary after the selections.

 \begin{table}[htbp]
  \begin{center}
   \caption{Background reduction summary and its efficiency.}
   \label{cut:table}
   \begin{tabular}{|l|r|r|r|r|}
    \hline
    Selection criteria & $H \to c\bar{c}$ & $H \to b\bar{b}$ & Higgs Bkg & SM Bkg\\
    \hline\hline
    Before the classification                & 2914 & 53480 & 23447 & 53333000 \\ \hline
    After the classification                 & 1693 & 29075 & 9198  & 20528900 \\ \hline
    $N_{charged} > 4$                        & 1238 & 22204 & 5721  & 3323060  \\ \hline
    $-\log(Y34)>2.7$                         & 1218 & 21869 & 5694  & 2635920  \\ \hline
    $thrust <0.95$                           & 1217 & 21858 & 5693  & 2584510  \\ \hline
    $|\cos{\theta_{thrust}}|<0.96$           & 1157 & 20831 & 5427  & 2295690  \\ \hline
    $105^{\circ} < \theta_{H} < 165^{\circ}$ & 1080 & 19393 & 4941  & 1908300  \\ \hline
    $ 70^{\circ} < \theta_{Z} < 160^{\circ}$ & 1028 & 18490 & 4705  & 1776150  \\ \hline
    $110 < M_{H_{fit}} < 140~{\rm GeV}$      &  982 & 17666 & 4411  & 1209100  \\ \hline
    $ 80 < M_{Z_{fit}} < 110~{\rm GeV}$      &  982 & 17665 & 4409  & 1206570  \\ \hline
    $E_{\gamma} < 20~{\rm GeV}$              &  895 & 16288 & 4063  & 1036990  \\ \hline\hline
    Efficiency after classification & 52.9\% $(\varepsilon_{cc})$ & 56.0\% $(\varepsilon_{bb})$ & 44.2\% & 5.0\%\\
    \hline
   \end{tabular}
  \end{center}
 \end{table}

 \section{Branching ratio measurement}

\begin{wrapfigure}{r}{0.6\columnwidth}
  \begin{center}
   \includegraphics[width=0.6\textwidth]{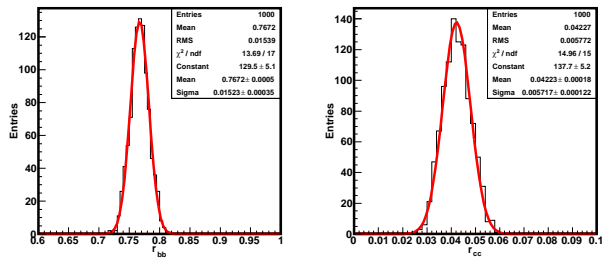}
   \caption{$r_{bb}$ and $r_{cc}$ distribution from the template fitting Toy-MC.}
   \label{rbb_rcc:fig}
  \end{center}
 \end{wrapfigure}

\begin{wrapfigure}{r}{0.35\columnwidth}
  \begin{center}
   \includegraphics[width=0.35\textwidth]{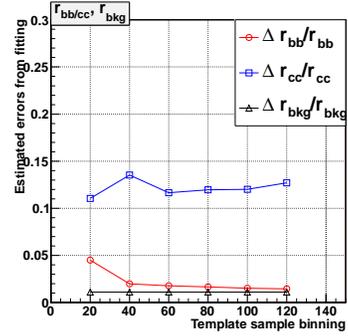}
   \caption{Binning dependence of the uncertainty of $r_{bb/cc}$ and $r_{bkg}$ after the smoothing.}
   \label{smoothing:fig}
  \end{center}
 \end{wrapfigure}

 Branching ratio ($BR$) of Higgs boson is related to the mass of the fermions and gauge bosons.
 After the background reduction, we evaluate the measurement accuracy of the branching ratios ($BR$).
 In this analysis, we evaluate the relative branching ratios of $H \to c\bar{c}$ to $H \to b\bar{b}$:
 $\displaystyle\frac{BR(H \to c\bar{c})}{BR(H\to b\bar{b})}$.
 In order to evaluate the measurement accuracy of $BR$,
 we apply the flavor-likeness template fitting~\cite{template} after the all event selections.
 For each flavor, neural-net training is performed using $Z \to q\bar{q}$ $Z$-pole ($\sqrt{s}=91.2~{\rm GeV}$) samples
 in LCFIVertex package.
 Here, $bc$-likeness is a $c$-likeness trained only with $Z \to b\bar{b}$ background.
 Each flavor-likeness for di-jet is defined as:
 $x$-likeness = $\displaystyle\frac{x_{1}x_{2}}{x_{1}x_{2}+(1-x_{1})(1-x_{2})}$,
 where $x_{1,2}$ are the neural-net trained output for $b$, $c$ and $bc$ flavor in each jet
 from the vertexing package of LCFIVertex~\cite{LCFI} in ilcsoft.
 \begin{figure}[htbp]
  \begin{center}
   \includegraphics[width=0.95\textwidth]{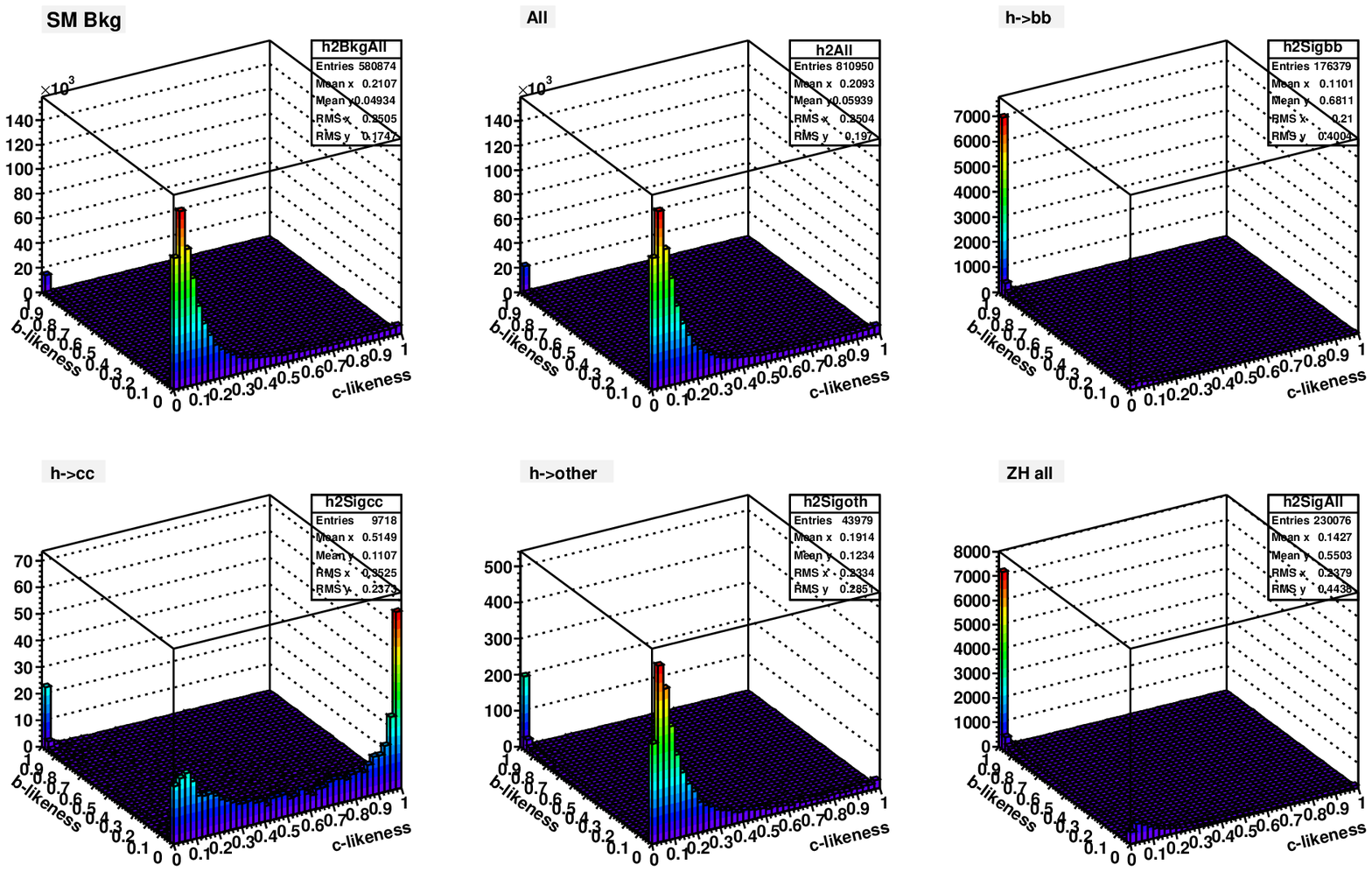}
   \caption{Image of the two-dimensional template sample histograms in $b$-likeness and $c$-likeness.}
   \label{2Dsample:fig}
  \end{center}
 \end{figure}
 The analysis procedure of the template fitting is as following:
 (1). Prepare the $b$, $c$ and $bc$ three dimensional flavor-likeness template samples for $H\to b\bar{b},~c\bar{c}, others$
 and $SM~{Bkg}$ decay modes;
 (2). Apply the Toy-MC template fitting test to evaluate the $r_{bb}$, $r_{cc}$,  $r_{others}$ and $r_{bkg}$,
 where $r_{bb}$, $r_{cc}$ are the number of entry ratios of $H \to b\bar{b},~c\bar{c}$
 after the selection cuts to the entry predicted from SM Higgs branching ratios,
 and $r_{others}$ is a ratio of another Higgs decay modes which is described as $r_{others}=1-r_{bb}-r_{cc}$,
 and $r_{bkg}$ is a normalized factor for the entries of SM background.
 (3). Estimate the relative branching fraction of $BR(H\to b\bar{b})/BR(H\to c\bar{c})$.
 In order to evaluate the fractions of Higgs decay mode in each sample,
 template fitting is applied with minimizing following $\chi^{2}$ value:
 \begin{equation} \label{chi2:eq}
  \chi^{2} =
   \displaystyle\sum_{i=1}^{n_{b}}\sum_{j=1}^{n_{c}}\sum_{k=1}^{n_{bc}}\frac{\left(N_{ijk}^{data}-\displaystyle\sum_{s=bb,cc,others}r_{s}\cdot\left(\frac{N^{ZH}}{N^{s}}\right){\cdot}N_{ijk}^{s}-r_{bkg} \cdot N_{ijk}^{bkg}\right)^{2}}{N_{ijk}^{all}},
 \end{equation}
 where $r_{s}$ represents the fitted parameters of $r_{bb}$, $r_{cc}$, $r_{others}$ and $r_{bkg}$.
 $N_{ijk}^{s}$ and $N_{ijk}^{bkg}$ are the number of expected entries in each 3D sample bin ($i,~j,~k$),
 since each histogram is separated into $n_{b}$, $n_{c}$ and $n_{bc}$ bins.
 $N_{ijk}^{data}$ are the number of simulated entries in each bin ($i,~j,~k$) by Toy-MC,
 which fluctuated with the Poisson distribution in each samples.\\
 In order to estimate the $r_{bb}$ and $r_{cc}$,
 template fitting is applied with fluctuating the data by Poisson distribution 
 and perform the 1,000 times Toy-MC analysis.
 Fitted $r_{bb}$ and $r_{cc}$ fractions are obtained from the distribution
 of the Toy-MC template fitted with Gaussian, as shown in Fig~\ref{rbb_rcc:fig}\\
 From the 1,000 times Toy-MC template fitting,
 we obtain the $r_{bb}$ and $r_{cc}$ as $0.767 \pm 0.002$, $0.422 \pm 0.006$,
 which reproduce the true signal fractions of $r_{bb}~(0.765)$ and $r_{cc}~(0.0422)$, respectively.\\
 From the limitation of the commonly reconstructed full simulation samples, especially for SM background, 
 number of entries in each template sample 3D histogram, which depends on the histogram binning,
 become a systematic uncertainty of the fitted parameters.
 In order to reduce the histogram binning dependence,
 we apply the histogram smoothing for template samples.
 Fig.~\ref{smoothing:fig} shows the binning dependence after the smoothing.
 \section{Results}
 The relative branching ratio in $H\to c\bar{c}$ to $H\to b\bar{b}$ is evaluated with following equation:
 \begin{equation}
  \label{relative_br:eq}
  \frac{BR(H\to c\bar{c})}{BR(H\to b\bar{b})}=\frac{r_{cc}/\varepsilon_{cc}}{r_{bb}/\varepsilon_{bb}},
 \end{equation}
 where the $\varepsilon_{bb/cc}$ are the efficiency of $H \to b\bar{b}/c\bar{c}$ events
 after the selections as shown in Table~\ref{cut:table}.
 From the Eq.~(\ref{relative_br:eq}), relative branching fraction is obtained as:\\
 $\displaystyle\frac{BR(H\to c\bar{c})}{BR(H\to b\bar{b})}=0.058 \pm 0.008$,
 which corresponds to the measurement accuracy of 13.68\%.

 \section{Conclusion}

 Measurement accuracy of Higgs branching ratio in Higgs hadronic decay mode $ZH \to q\bar{q}H$ in ILC experiment
 is evaluated at the $\sqrt{s}=250~{\rm GeV}$
 with assuming the Higgs mass of $M_{H}=120~{\rm GeV}$ and the integrated luminosity of ${\cal L}=250~{\rm fb}^{-1}$.
 From the template fitting analysis,
 measurement accuracy of the relative branching ratio
 of $H\to c\bar{c}$ to $H\to b\bar{b}$ is evaluated as 13.68\%. 
 
 \section*{Acknowledgment}
 
 I would like to thank to everyone who join the ILC physics WG subgroup~\cite{physwg}
 for useful discussion of this work
 and to ILD optimization group members who maintain the software and MC samples.



{\footnotesize
 \begin{thebibliography}{99}
  \bibitem{RDR} ILC Reference Design Report (RDR) http://www.linearcollider.org/rdr/
  \bibitem{LEPEWWG}
  J.~Alcaraz  [ALEPH Collaboration and CDF Collaboration and D0 Collaboration and  an],
  arXiv:0911.2604 [hep-ex].
  \bibitem{CDF}
  C.~Anastasiou, G.~Dissertori, M.~Grazzini, F.~Stockli and B.~R.~Webber,
  JHEP {\bf 0908}, 099 (2009)
  [arXiv:0905.3529 [hep-ph]].
  \bibitem{Geant4}   GEANT4 Collaboration: S Agostinelli et al, Nucl. Instrum. Methods A506, 250 (2003).
  \bibitem{ilcsoft}  http://ilcsoft.desy.de/portal/
  \bibitem{Root}     http://root.cern.ch/
  \bibitem{SimTools} http://acfahep.kek.jp/subg/sim/simtools/
  \bibitem{LCFI}     Nuclear Instruments and Methods in Physics Research Section A, Volume 610, Issue 2, p. 573-589.
  \bibitem{template} T.~Kuhl and K.~Desch, LC-PHSM-2007-001
  \bibitem{physwg}   http://www-jlc.kek.jp/subg/physics/ilcphys/
\end{thebibliography}
}
\end{document}